\documentclass[12pt]{article}
\usepackage{amsfonts}
\usepackage{amsmath}
\usepackage{amssymb}
\usepackage{amsthm}
\usepackage{graphicx}
\usepackage{latexsym}
\usepackage{xy}
\usepackage{verbatim}
\usepackage{bm}
\usepackage{hyperref}

\parskip=\baselineskip
\begin{document}

\begin{titlepage}
\begin{flushright}
\end{flushright}
\vskip 1.5in
\begin{center}
{\bf\Large{Comment on ``Cosmological Topological Massive Gravitons
and Photons"}} \vskip 0.5in {Wei Li, Wei Song\footnote{Permanent
address: \textit{Institute of Theoretical Physics Academia Sinica,
Beijing 100080, China.}} and Andrew Strominger}

\vskip 0.3in
{\small{\textit{ Center for the Fundamental Laws of Nature\\
Jefferson Physical Laboratory, Harvard University, Cambridge MA
02138, USA}}}

\end{center}
\vskip 0.5in

\baselineskip 16pt
\date{}

\begin{abstract}
In a recent paper \cite{lss} it was shown that all global energy
eigenstates of asymptotically $AdS_3$   chiral gravity have
nonnegative energy at the linearized level. This result was
questioned \cite{cdww}  by Carlip, Deser, Waldron and Wise (CDWW),
who work on the Poincare patch. They exhibit a linearized solution
of chiral gravity and claim that it has negative energy and is
smooth at the boundary. We show that the  solution of CDWW is smooth
only on that part of the boundary of $AdS_3$ included in the
Poincare patch. Extended to global $AdS_3$,  it is divergent  at the
boundary point not included in the Poincare patch. Hence it is
consistent with the results of \cite{lss}.
\end{abstract}
\end{titlepage}
\vfill\eject

We consider the theory of
TMG (topologically massive gravity) with a negative
cosmological constant. This is described by   \cite{Deser:1981wh}\cite{Deser:1982vy}\cite{DeserTekin1}\footnote{We have chosen here the standard sign in front of the Einstein-Hilbert  action so that BTZ black holes have positive energy for large $\mu$.This contrasts with most of the literature which chooses the opposite sign in order that massive gravitons have
 positive energy (for large $\mu$).}
\begin{equation}
I= \frac{1}{16\pi G}\int d^3x \sqrt{-g}
(R+\frac{2}{\ell^2})+\frac{1}{16\pi G\mu}I_{CS}
\end{equation}
where
 $I_{cs}$ is the Chern-Simons term
\begin{equation}
I_{cs}=-\frac{1}{2}\int
d^3x\sqrt{-g}\epsilon^{\lambda\mu\nu}\Gamma^\rho_{\lambda\sigma}
[\partial_\mu\Gamma^\sigma_{\rho\nu}+\frac{2}{3}\Gamma^\sigma_{\mu\tau}\Gamma^\tau_{\nu\rho}].
\end{equation}
 TMG has an
$AdS_3$ vacuum solution:
\begin{equation}\label{AdS3metric}
ds^2=\ell^2(-
\cosh^2{\rho}d\tau^2+\sinh^2{\rho}d\phi^2+d\rho^2)
\end{equation}
For generic signs and values of the parameters TMG will have
solutions which are asymptotically $AdS_3$ (by which we always mean obeying the precise boundary conditions given in
Brown and Henneaux \cite{brown} and discussed in the context of  TMG recently in \cite{bhtmg}) but have negative energies. This limits the physical interest of the generic TMG theory. However in
 \cite{lss} it was pointed out that for the case $\mu\ell=1$ of  so-called chiral gravity all known linearized and nonlinear asymptotically $AdS_3$ solutions have nonnegative energy. It was further conjectured that  both the classical and quantum theories of chiral gravity are consistent and stable. The terminology chiral gravity was adopted because
 the central charge of the dual CFT, if it exists,  must be  \cite{brown}\cite{klarsen}\cite{kraus}
 $(c_L,c_R)=\frac{3\ell}{2G}(1-\frac{1}{\mu\ell}, 1+\frac{1}{\mu\ell})$ and hence is purely chiral  at $\mu\ell=1$.

Solutions of the linearized equations which are energy eigenstates (where energy is defined as in \cite{DeserTekin2}\cite{Olmez:2005by}using  the global time $\tau$) fall into $SL(2,R)_L\times SL(2,R)_R$ representations which can be characterized by the $(L_0,\bar L_0)$ eigenvalues $(h_L,h_R)$ of the highest weight states.
In \cite{lss} it was shown that there are three representations labelled by the highest weights $(h_L,h_R)$
\begin{equation}(2,0),~~~~(0,2),\end{equation}
corresponding to massless  left and right moving boundary gravitons
and
\begin{equation}\label{mg} (\frac{3}{2}+\frac{\mu\ell}{2}, -\frac{1}{2}+\frac{\mu\ell}{2}),
\end{equation}
corresponding to the so-called massive gravitons. For $\mu\ell > 1$,  despite the
positivity of the weights, the massive gravitons have negative energy
because the kinetic term has an overall negative prefactor of
$(\frac{1}{\mu\ell}-1)$.

It was pointed out \cite{lss} that for chiral gravity, the Compton wavelength of the massive graviton becomes of order the $AdS_3$ radius, the weights and the wavefunctions of the massless left-moving and massive gravitons degenerate and the energies of both becomes zero.
Hence all known asymptotically $AdS_3$ solutions have nonnegative
energy in this special case.

CDWW claim there are asymptotically $AdS_3$ solutions of TMG with negative energy even at the chiral point.\footnote{Actually to be more precise CDWW use a negative Newton's constant so the energy is positive, but with our sign choice it is negative.}  They work in Poincare coordinates for which the line element is  (CDWW equation (35))
\begin{equation}\label{eomz}ds^2=\frac{2dx^+dx^-+dz^2}{z^2}.
\end{equation}They find a continuum (rather than a discrete set) of solutions of the linearized wave equations, work out the linearized Einstein tensor ${\cal H}_{\rho\sigma}$
(CDWW equation (44)). They find that e.g. ${\cal H}_{--}$ is given by
(CDWW equation (48) and page 12)
\begin{equation}\label{meom}
{\cal H}_{--}\sim z J_3(z) \to z^4,~~~z\to 0
\end{equation}
where $J$ is a Bessel function. This indeed vanishes
at the boundary $z=0$ of the Poincare patch. However this does not cover all of the boundary points of global $AdS_3$. The missing boundary point can be approached by fixing Poincare time $x^+-x^-=constant$
and taking $z\to \infty$. From Bessel function asymptotics it is then easily seen that the curvature diverges as
\begin{equation}\label{meom}
{\cal H}_{--}\sim \sqrt{z}\cos z,~~~
,~~~z\to \infty
\end{equation}
In a local orthonormal frame the divergence is faster by a factor of
$z^2$. A simple analogy is the function $f(z)=z$  in the upper half
complex $z$ plane. It vanishes everywhere on the boundary $Im ~z=0$.
However the boundary of the Poincare disc includes the point $Im
~z=\infty$ where $f$ diverges.

In conclusion, the curvature tensor of the CDWW negative-energy
massive graviton solutions of chiral gravity diverges on the
boundary of global $AdS_3$. This is fully consistent with the
observation of \cite{lss} that all asymptotically $AdS_3$ $(L_0,\bar
L_0)$ eigensolutions of linearized chiral gravity have nonnegative
energy.

\section*{Acknowledgements}
We thank S. Carlip, D. Grumiller and A. Maloney  for helpful discussions. The work is supported by
DOE grant DE-FG02-91ER40654. WS is also supported by CSC. WS would
like to thank the hospitality of the physics department of Harvard
University.


\end{document}